# Optical elastic scattering for early label-free identification of clinical pathogens

Valentin Genuer*[a], Olivier Gal[b], Jérémy Méteau[a], Pierre Marcoux[a], Emmanuelle Schultz[a], Éric Lacot[c], Max Maurin[d], Jean-Marc Dinten[a]

[a] Université Grenoble-alpes, F-38000 Grenoble, France
CEA, LETI, Minatec-Campus, 17 avenue des Martyrs, 38054 Grenoble Cedex 9, France
[b] CEA, LIST, Gif-sur-Yvette, F-91191, France
[c] Université Grenoble-alpes, F-38000 Grenoble, France
Liphy, Laboratoire interdisciplinaire de physique, 140 Rue de la Physique, 38402 Saint-Martin-d'Hères, France
[d] Laboratoire de Bactériologie, Département des agents infectieux, Institut de biologie et pathologie, CHU de Grenoble, CS 10217, 38043 Grenoble Cedex 9, France

**ABSTRACT**

We report here on the ability of elastic light scattering in discriminating Gram+, Gram- and yeasts at an early stage of growth (6h). Our technique is non-invasive, low cost and does require neither skilled operators nor reagents. Therefore it is compatible with automation. It is based on the analysis of the scattering pattern (*scatterogram*) generated by a bacterial microcolony growing on agar, when placed in the path of a laser beam. Measurements are directly performed on closed Petri dishes.

The characteristic features of a given scatterogram are first computed by projecting the pattern onto the Zernike orthogonal basis. Then the obtained data are compared to a database so that machine learning can yield identification result. A 10-fold cross-validation was performed on a database over 8 species (15 strains, 1906 scatterograms), at 6h of incubation. It yielded a 94% correct classification rate between Gram+, Gram- and yeasts. Results can be improved by using a more relevant function basis for projections, such as Fourier-Bessel functions. A fully integrated instrument has been installed at the Grenoble hospital's laboratory of bacteriology and a validation campaign has been started for the early screening of SA and MRSA (S*taphylococcus aureus*, methicillin-resistant *S. aureus*) carriers.

Up to now, all the published studies about elastic scattering were performed in a forward mode, which is restricted to transparent media. However, in clinical diagnostics, most of media are opaque, such as blood-supplemented agar. That is why we propose a novel scheme capable of collecting back-scattered light which provides comparable results.

**Keywords**: Optical elastic scattering, bacteria identification, bacterial colony, early screening, optical elastic backward-scattering, SVM, Fresnel diffraction patterns, MRSA.

# 1. INTRODUCTION

Improved, fast and accurate detection and diagnostic technologies are needed in order to address present and future needs for rapid identification of contaminant or pathogens such as bacteria and yeasts for both industrial microbiological control and clinical diagnostics [1].

Although traditional culturing methods employing classical microbiological techniques are considered as gold standards because they provide high accuracy rates, they suffer from lengthy process due to the need of multiple preparation steps; taking up to 3-7 days to obtain negative results and additional 2-3 days to confirm a positive result by using biochemical or molecular techniques [2]. Thus, since a decade, there is a growing interest for alternative methods based on light scattering which potentially have considerable advantage over the standards because they allow fast and non-invasive direct on agar analysis of bacterial colonies. The overall process of those techniques based on light scattering starts by noting that differences in refractive index, biomass, characteristics sizes and internal arrangement of bacteria inside the colonies leads to different forward scattering patterns when shooting it with a laser beam. Then with the help from learning algorithms it is possible to build a model based on a database containing numerous scattering patterns corresponding to different microorganisms and thus to provide discrimination results between them [3-5].

Advanced Bioimaging System was the first to launch in 2009 his BARDOT: a bacterial rapid detection system using optical light scattering developed at Purdue University which is able to provide precise identification of major foodborne pathogens. Around 90-99% for a discrimination at genus or species level for *Listeria, Salmonella, Escherichia* or *Staphylococcus* [6,7]. However his analysis throughput is limited by the incubation time needed to let the bacterial colonies reach a certain size, typically around 2 millimeters. Recently, a study by P.R. Marcoux *et al* reported some results about the development of a new kind of system using elastic forward scattering which provide good discrimination rates at species and strain level on bacterial and yeasts microcolonies at an early stage of growth. For example an 87% recognition rate for a strain level discrimination of four strains of *Escherichia coli* after 6h of incubation at 37°C was reached. The novel setup they proposed combined two optical paths in order to acquire images of microcolonies both in direct space (morphology insight) and in reciprocal space (diffraction pattern). Furthermore they used a specific laser beam profile to match the variability in size of the different microorganisms after 6h of growth (between 20µm and 250µm) [8].

The present study inspired by these recent developments reports on the improvements we brought both to the optical design and the image analysis so as to increase classification performances and also to address clinical applications involving non-transparent growth culture media.

It first deals with the optical design and the results obtained in discriminating Gram+, Gram- and yeasts at 6 hours with our forward-scattering system; the screening of SA and MRSA carriers will also be explained. Then it details the improvement that can be achieved by working on the analysis process, especially by changing the orthogonal basis on which the images are projected to build the features vectors.

Yet, the technique is still restricted to transparent culture media due to transmission acquisition geometry, and is consequently not adapted to coloured media such as blood-supplemented agar media. Those opaque media are of common use in clinical microbiology and up to now, there is not any system based on elastic scattering that can provide comprehensive scattering patterns from microcolonies growing on such culture media. A study, carried out in 2004 by S. Guo showed that forward scattering was more promising than back-scattering method because they could not build a system capable of providing patterns of good quality [9].

This is why we propose a novel system that could enlarge the application domain of elastic light scattering methods to the use of opaque culture media by providing an optical architecture capable of overcoming this challenge and thus collecting comprehensive patterns from light back-scattered by microcolonies directly on their commercial Petri dishes. We show here the solutions we provide to build such a system and we also show the results obtained after testing its discrimination performance at the species or strain level on different bacterial and yeasts species. For the testing, a database was acquired using a common opaque culture media (blood-supplemented agar COS) used in clinical microbiology to make fastidious organisms grow.

## 2. MATERIALS AND METHODS

### 2.1 Sample preparation

Bacterial strains were obtained from KwikStik (Microbiologics, St. Cloud, MN) lyophilized ATCC reference strains. Starting from a culture on COS or TSA (bioMérieux) after 24h of incubation (37°C), a 0.5 McF suspension (Densimat, bioMérieux) was prepared in API NaCl 0.85 % Medium (bioMérieux). This 0.5 McF was then immediately diluted at 1/120 in NaCl 0.85 %. 50µL of this diluted suspension were finally plated on agar medium (90mm petri dish, bioMérieux), either on ChromID S. aureus or TSA, for 6h of additional incubation and scattering experiments.

### 2.2 Instruments

The scatterograms were acquired using the forward-scattering setup depicted in Figure 1. A 532 nm laser (Spectra Physics Excelsior 532-50-CDRH) produces a coherent collimated light beam with a power decreased to less than 1mW thanks to an optical density. A pair of achromatic lenses is used to give this beam a particular shape, allowing us to reach a range of laser spot sizes from 10µm to 300µm by moving the sample along the beam path. A first video camera (Pixelfly QE) is used to acquire the forward scatterograms resulting from the interaction between the laser beam and the targeted microcolony. A second video camera coupled with achromatic lenses is used to get a microscope-like image of the sample. This bright field image is obtained when the sample is put at the focus point of the 2-lenses-system and illuminated by a ring LED placed above it. This configuration is convenient because it provides both a direct image of the targeted microcolony and its corresponding forward scatterogram. The alignment of the sample with the laser beam, and the adjustment of the laser spot size on the sample are achieved using horizontal and vertical translation stages (PI Micos VT75).

Recently, this setup has been integrated in a transportable instrument, the so-called MICRODIFF system, depicted in Figure 2. It is currently being tested at the Grenoble hospital's laboratory of bacteriology for the screening of the SA and MRSA (*S. aureus* and methicillin-resistant *S. aureus*) carriers. Since the *Staphylococci* that were used for the database grow slowly, we had to replace the lens above the sample by a microscope objective (10x, NA: 0.25) so as to get a laser spot size that could reach the microcolonies sizes after a 6h incubation time. At this time of growth, indeed, some strains were only around 4 µm long.

To enlarge the application domain of elastic scattering to clinical diagnostic purpose needing Petri dishes filled with opaque culture medium, we build another system, depicted in Figure 3, able to acquire scatterograms backward reflected by the sample. The same laser source coupled with achromatic lenses is used to produce a light beam with a particular shape; the same principle is used to vary the laser spot size on the sample: we move the sample across the beam path. There is a direct imaging system too composed of a CMOS video camera and a couple of achromatic lenses (orange path in Figure 3). The focus is obtained by placing the sample at the focal plane of the lens near it. To acquire scatterograms backward reflected we used a quarter-wave plate coupled with a polarizing beam splitter (blue path in Figure 3). The light issued from the laser source is linearly polarized and is first reflected toward the sample thanks to the polarizing beam splitter (Thorlabs, CM1-PBS251). As it passes through the quarter-wave plate, it becomes rotationally polarized. The reflection on the sample inverts the direction of polarization and as the light passes back through the quarter-wave plate it becomes linearly polarized but perpendicularly to the initial one. Thus it is now transmitted through the polarizing beam splitter and then collected by an achromatic lens and finally acquired on a CMOS video camera. This system could reach a spot size range from 30 µm to 100 µm and had a numerical aperture around 0.15.

This setup has also recently been upgraded by adding a microscope objective (20x magnification, NA: 0.45) to increase the numerical aperture and thus get more diffraction rings and also get a microscope image with a greater magnification.

The main drawback of adding optical elements between the polarizing beam splitter and the sample is getting specular reflections that saturate the CMOS sensor. This is managed by tilting some elements, such as the quarter-wave plate while adjusting the polarization with the quarter-wave plate.

## 2.3 Image Analysis

*Zernike moments invariants*

As a first choice for features-extraction from images, the use of Zernike moments invariants was quite relevant because of the typically circularly shaped scattering patterns. Basically, those scatterograms were initially centered in the middle and histograms were equalized. Then Zernike moments invariants were calculated up to the 20$^{th}$ order.

Zernike moments are computed from a complete orthogonal basis set of polynomials defined on the unit disk [10]. The expression of these Zernike orthogonal basis functions in polar coordinates is classically:

$$V_{nm}(r,\theta) = R_{nm}(r)\exp(jn\theta),$$

where $(r,\theta)$ are defined on the unit disk, *n* is a non-negative integer, |*m*| is less or equal to *n* and *n*-|*m*| must be even. $R_{nm}$ is the Zernike radial polynomial, defined in detail in [11] for example.

Zernike moments are calculated after projecting the images on this orthogonal basis. Thus if we call *f* the digital image function, the Zernike moment of radial order *n* with azimuthal repetition *m* is expressed as:

$$Z_{nm}(r,\theta) = \frac{n+1}{\pi}\sum_x\sum_y f(x,y)V_{nm}^*(r,\theta)dxdy, \ x^2 + y^2 \leq 1.$$

As rotation-invariant features we took the magnitudes of Zernike moments as they remain unchanged under rotation. Computing these moments up to the 20$^{th}$ order gave us features-vector of 121 components per image which then were used to train a supervised machine learning algorithm. The main idea was to build a model that correctly mapped the features-vector to the corresponding sample (*e.g.* bacterial strain). The process, previously detailed [8], follows three main steps: first the acquisition of a large database composed of hundreds of images per class; then the training of the learning algorithm, using an SMO (Sequential Minimal Optimization) which is a particular algorithm used to train SVM (Support Vector Machines) while reducing calculation time; finally a cross-validation to test the eventual classifier errors and summarize them in a confusion matrix. The WEKA data mining software was used for this last task.

A first observation about Zernike radial polynomials properties is that at the order *n* we can at most obtain *n/4* complete oscillations. This limits at *n/4* the number of rings in scatterograms that can be caught by the projection. For example (Figure 4a), at the order *n*=20 we can hope to retrieve at most 5 rings. This is too few compared to the actual number of rings present in the scatterograms we observed, and increasing the order would have drastically increased the features-vector size. Moreover we can note that intensity always reaches his maximum for *r*=1 (*i.e.* at the unit circle boundaries) while we observed on our images that intensity was rapidly decreasing with *r*. Thus it could be problematic in terms of expansion convergence. This is why we decided to experiment another projection basis.

*Fourier-Bessel moments invariants*

Fourier-Bessel moments are computed from a basis of functions based on the Bessel function of the first kind. The functions:

$$J_n(\alpha_{nm}r)e^{\pm jn\theta},$$

where $n \geq 0$, $m \geq 1$ and $\alpha_{nm}$ is the $m^{th}$ root of the $J_n$ Bessel function, form an orthogonal basis on the unit disk $r \leq 1$ [12]. Calling $f$ the image function, Fourier-Bessel moments are defined in polar coordinates as follows:

$$F_{nm} = \frac{1}{2\pi c_m} \int_0^{2\pi} \int_0^1 f(r,\theta) J_n(\alpha_{nm}r) e^{-jn\theta} r\, dr\, d\theta,$$

where $m \geq 1$, $n = 0, \pm 1, \pm 2, \ldots$ is the moment order, $c_m$ is a normalization constant and $\alpha_{nm}$ is the $m^{th}$ root of the $J_n$ Bessel function. As we did with the Zernike moments, we took the magnitudes of the Fourier-Bessel moments to build our rotation invariants.

It was previously shown [13] that Fourier-Bessel moments take advantages of the zeros repartition of the Fourier-Bessel radial function $J_n(\alpha_{nm}r)$, which are uniformly distributed over the interval $0 \leq r \leq 1$ (Figure 4b for $n = 20$) unlike the zeros of the Zernike radial polynomials which are mainly located at the unit circle boundaries (Figure 4a).

## 3. RESULTS AND DISCUSSION

### 3.1 Gram+/Gram-/Yeasts discrimination

Figure 5 presents some scatterograms of the first database acquired in this work. 1906 scatterograms were acquired over the course of 3 months on Gram+, Gram- and yeasts (more than 120 scatterograms per strain) for a total of 15 different strains cultured in the conditions described above. Petri dishes lids remained closed during the whole acquisition process. The features-vectors were computed using 121 Zernike moments and the SMO was trained. The 10-fold cross-validation showed a global correct classification rate of 75% down to the strain level (details not shown). A more relevant classification rate of around 95% was obtained for the discrimination between Gram+, Gram- and yeasts. The confusion matrix is detailed in Table 1.

### 3.2 Early screening of SA/MRSA carriers at 6h

A validation campaign has been started at the Grenoble's hospital laboratory of bacteriology on the early screening of *Staphylococcus aureus* (SA) and methicillin-resistant *S. aureus* (MRSA) carriers. The acquisitions were made at 6h of growth with the fully integrated instrument version of the forward-scattering system MICRODIFF. A specific plate, the ChromID *S. aureus* was used. In fact this is a bi-plate filled with a specific medium for the growth of *Staphylococci*. One side contains a growth medium enriched with methicillin to detect resistant *Staphylococci* and the other side is only filled with the growth medium. The database, which is still in acquisition, is restrained to clinical relevant *Staphylococci* growing on ChromID and some interfering strains (*i.e.* non-*Staphylococci* strains which are growing on ChromID, *e.g.* some gram- like *Stenotrophomonas matlophilia* and yeast like *Candida tropicalis*). At this time we are able to get an average correct classification rate of 75.5% ± 0.2% (average on ten 10-fold cross-validations) between *S. aureus* and other *Staphylococci* (using 121 Zernike moments). This is a promising result at 6h of growth. Nevertheless this clinical application shows the difficulty of the exercise in terms of bacterial microcolony size. As we can see in Figure 6, scatterograms acquired on *Staphylococci* after only 6h of growth are quite similar and simple. This is principally due to the small size of the microcolonies and also the low biomass, since those strains are growing slowly.

## 3.3 Validation of the backward scattering system on four *E. coli* strains

We decided to validate our backward scattering system by acquiring a database on four *Escherichia coli* strains commonly associated with human infections (ATCC 25922, 35421, 11775 and 8739). We used the same four strains as P. Marcoux *et al* [8] on their forward scattering system to test the strain-level discrimination. Commercial COS Petri dishes (Columbia agar +5% sheep blood) were inoculated and incubated during 6h at 37°C. Then the acquisition was performed using the instrument described above. The direct space imaging unit for fast colony localization and targeting combined with short acquisition time (around 10 ms) allowed to acquire around 100 scatterograms per strain in a week, for a database of 400 images. Figure 7 shows the scatterogram acquired in reflection on a bacterial microcolony. It also displays the microscope image of the microcolony compared to the one observed with the bright field unit of the setup.

To rigorously compare the performances of the backward scattering system to the forward scattering system ones, we used the same features-vectors (121 Zernike invariants moments) to describe the images. Then the same SMO learning algorithm coupled with a 10-fold cross-validation was used. The confusion matrix showed an overall correct classification rate of 87% (Table 2), which is close to the 90% obtained with the forward scattering system and thus validates the new instrument. Promising results were also obtained on the discrimination of three *Candida* yeasts species (ATCC 14053, 2001 and 18303) where we could reach an overall classification rate of 94% (details not shown).

## 3.4 Analysis improvements with Fourier-Bessel moments

We made two main observations about the use of Fourier-Bessel moments: first we looked at the accuracy of the projection by comparing Zernike and Fourier-Bessel projections on different scatterograms and then we tried to evaluate how the number of moments used for the features-extraction influenced the classification rates.
Figure 8 shows the comparison between the projections of a scatterogram on Zernike basis and on Fourier-Bessel basis. Once the projections were made, we computed the reconstruction and calculate the relative error (to the original image). Approximately the same amount of coefficients was used: 121 for the Zernike projection (maximal radial order: 20, maximal azimuthal order: 20) and 120 for the Fourier-Bessel one (maximal radial order: 24, maximal azimuthal order: 4). It showed a better reconstruction accuracy for Fourier-Bessel, 37%, against 66% for Zernike with approximately the same number of moments.

We also showed that increasing the number of moments gives a more faithful reconstruction and thus a lower reconstruction error. Figure 10 illustrates that at equal number of moments, using Zernike basis leads to higher reconstruction error rates. On this example we reached a plateau at around 30% reconstruction error with the Fourier-Bessel basis. This level was reached from 160 Fourier-Bessel moments. It could be due to the noise level and the presence of local defects in the image. The Zernike reconstruction error seemed to decrease steadily but was always far higher than the Fourier-Bessel.

We used the Fourier-Bessel moments on the 4-strains *E-coli* database acquired with the backward system. We reached an average classification rate of 93.0% ± 0.7% (average over ten 10-fold cross-validations; confusion matrix given in Table 3). Compared to the 87% obtained when using Zernike moments, it corresponds to a decrease of the classification error rate by a factor 1.9.

## 4. CONCLUSION

The MICRODIFF instrument was employed to perform fast identification of bacterial pathogens down to the strain level. Several scatterograms databases were acquired at an early stage of growth in a forward or backward scattering mode. A correct classification rate of 94% was achieved on the Gram+/Gram-/Yeasts discrimination using the forward scattering system and a SVM classifier in a 10-fold cross-validation.

The backward scattering system allowed us to reach 87% on the classification of 4 *E. coli* strains. This has validated the optical reflection design for the identification of pathogens directly on opaque agar media. A novel setup is now being developed to combine both forward and backward scattering acquisitions and also the direct bright field imaging unit.

We also showed we could improve these results by using Fourier-Bessel moments instead of Zernike moments for the features extraction from scattering patterns. The correct classification on the 4 *E. coli* strains database rate was increased to 93%, corresponding to a reduction of the classification error rate by a factor 1.9. This shows that decreasing the reconstruction error and thus finding the best way to faithfully describe the scatterograms leads to higher classification rates. Further work is needed to more fully understand how much we can still gain in classification performance and also to determine whether the classification process relies on the whole features vector or only on few components.

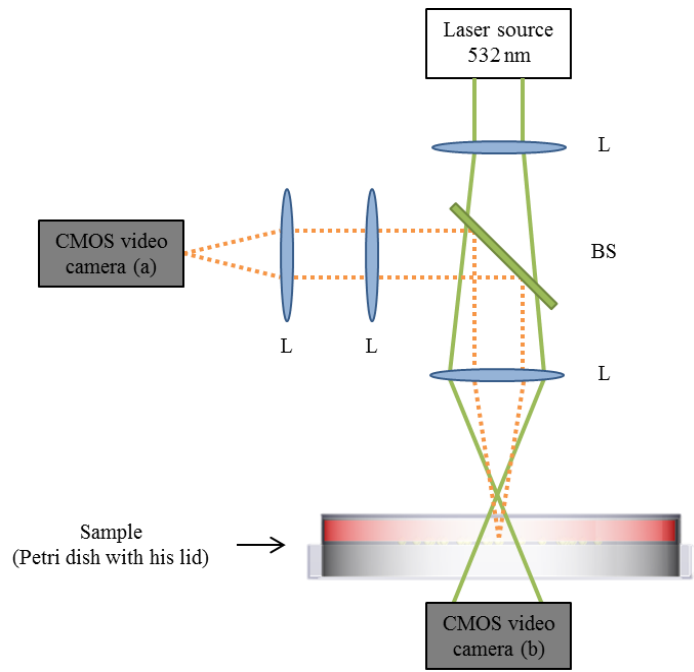

Figure 1 Schematic of the forward-scattering experimental setup for the acquisition of scatterograms on transparent growth media. Abbreviations: L: achromatic lens, BS: beam splitter, RL: ring LED. CMOS sensors: (a) for the direct space imaging, (b) for the acquisition of the scatterograms.

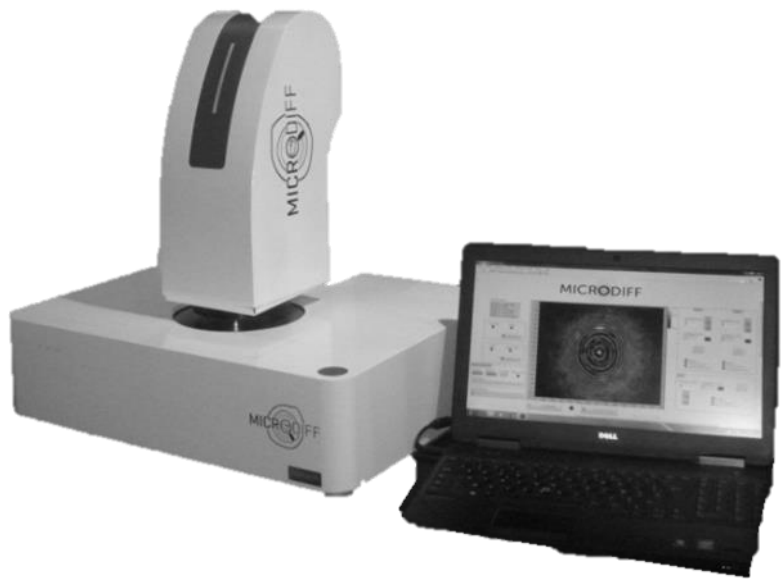

Figure 2 Photograph of the integrated MICRODIFF instrument for fast identification of clinical pathogens using elastic light scattering. Dimensions: 63x21x21 cm3, 10kg.

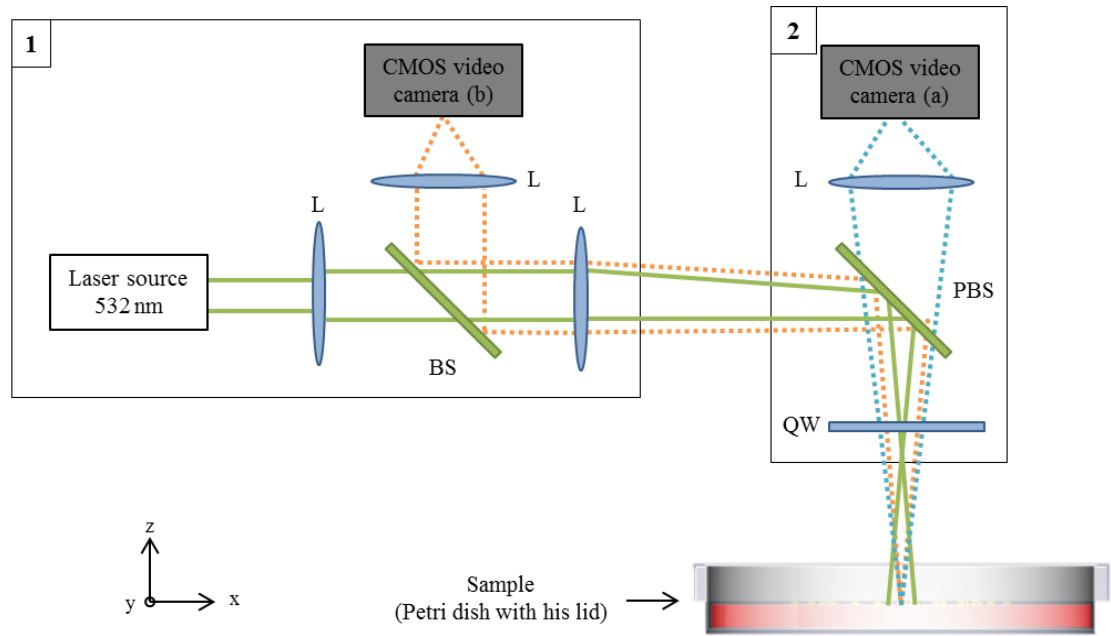

Figure 3 Schematic of the backward-scattering system. Abbreviations: L: achromatic lens, BS: beam splitter, PBS: polarizing beam splitter, RL: ring LED. CMOS sensors: (b) for the direct space imaging, (a) for the acquisition of the scatterograms.

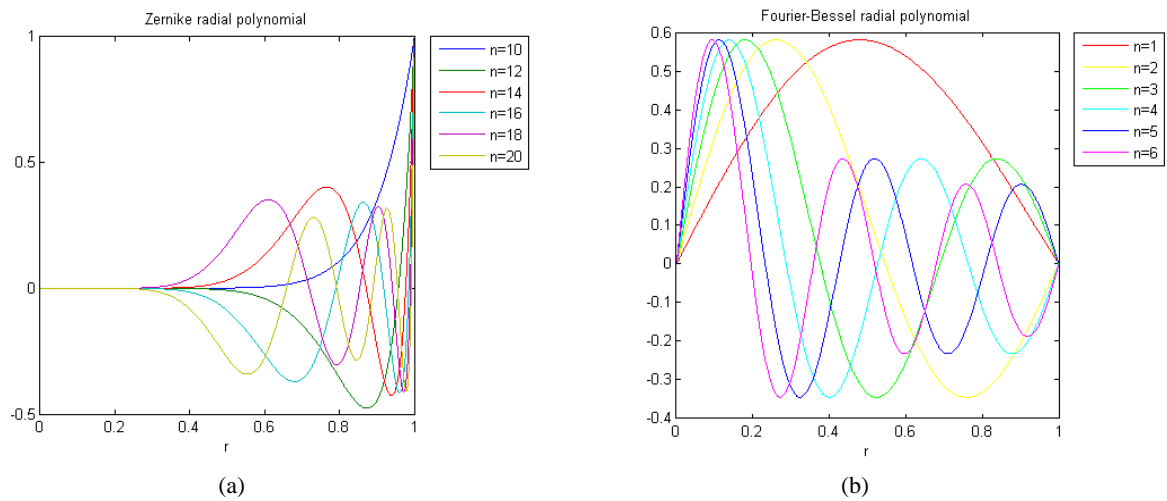

(a)          (b)

Figure 4 (a) Zernike radial polynomials $R_{nm}(r)$ for m=10. (b) Radial functions $J_1(\alpha_n r)$ for the Fourier-Bessel projections with azimuthal order 1. We can note the more convenient zeros distribution of the Fourier-Bessel radial functions.

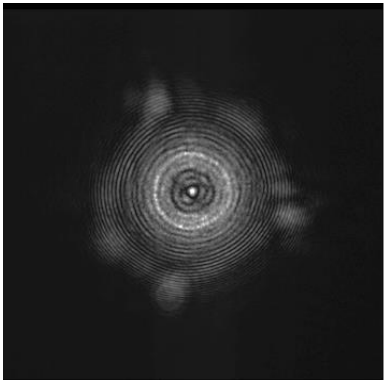 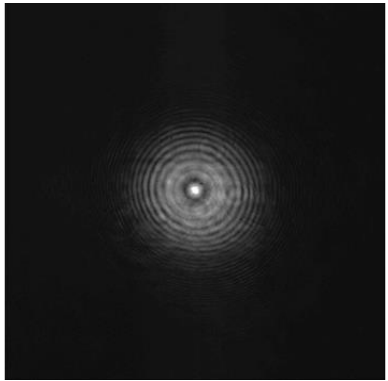 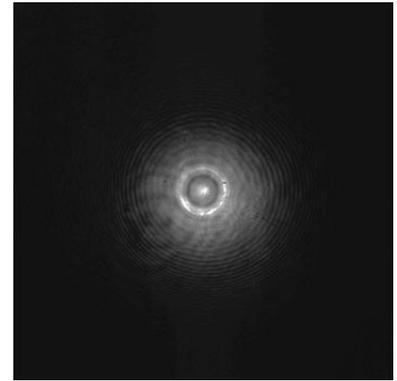

ATCC 25922  
*Escherichia coli*

ATCC 18804  
*Candida albicans*

ATCC 49741  
*Staphylococcus epidermidis*

Figure 5 Scatterograms of microcolonies acquired on the forward-scattering setup after 6h of growth at 37°C on TSA.

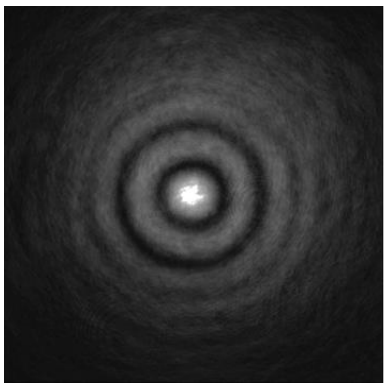 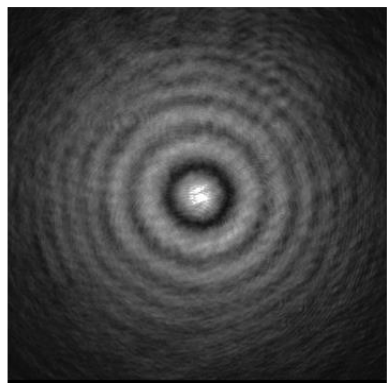 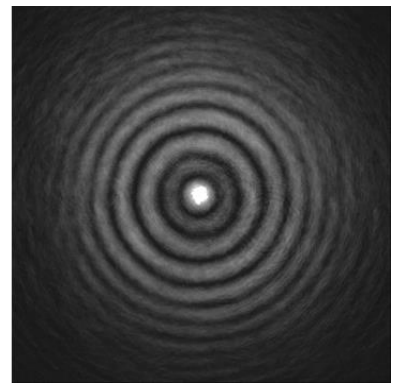

ATCC BAA-44  
*Staphylococcus aureus subsp. aureus*

ATCC 27851  
*Staphylococcus simulans*

ATCC 29970  
*Staphylococcus haemolyticus*

Figure 6 Scatterograms of microcolonies acquired after 6h of growth at 37°C (on ChromID *S. aureus*) with the forward scattering prototype at the Grenoble's hospital laboratory of bacteriology during the validation campaign on the early screening of SA/MRSA carriers.

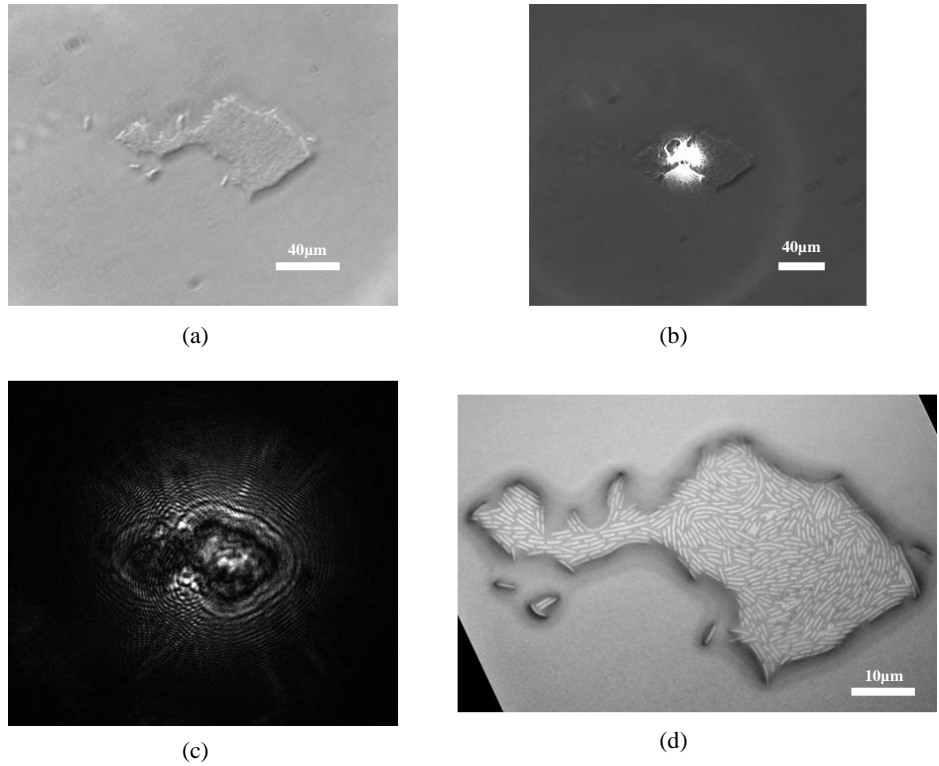

Figure 7 Images of an *E. coli* GFP microcolony acquired on the backward-scattering setup at 6h of growth on LB + ampicillin. (a) Direct space imaging. (b) Direct space imaging plus laser spot targeting the micro-colony. (c) Scatterogram acquired on the setup. (d) Microscope image with a 50x objective, AxioVision Zeiss.

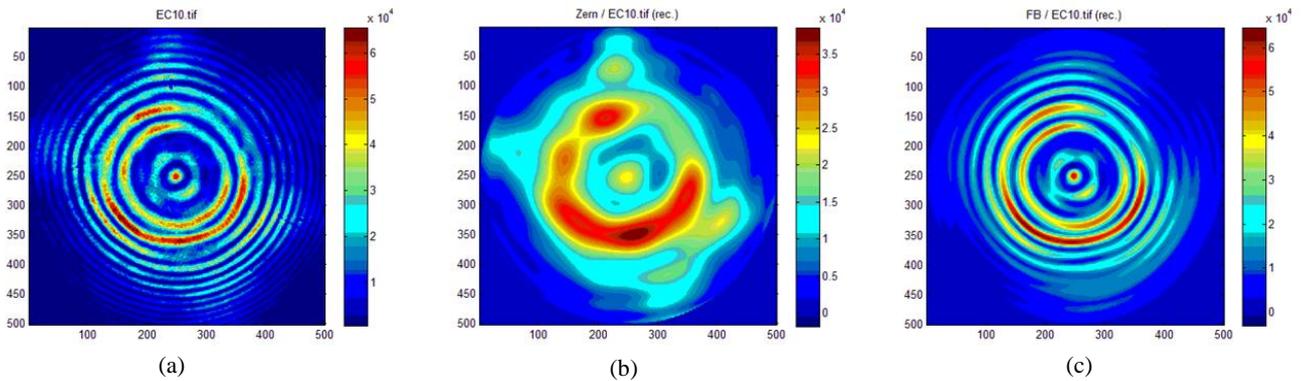

Figure 8 Comparison of the image reconstructions obtained using Zernike and Fourier-Bessel expansions. (a) Scatterogram of an *E. coli* ATCC 25922 microcolony acquired with the forward-scattering system MICRODIFF after 6h at 37°C. (b) Reconstruction after projection on the 121 first Zernike moments (maximal radial order: 20, maximal azimuthal order: 20). Quadratic error: 66.3%. (c) Reconstruction from the projection on 120 Fourier-Bessel moments (maximal radial order: 24, maximal azimuthal order: 4). Quadratic error: 37.7%

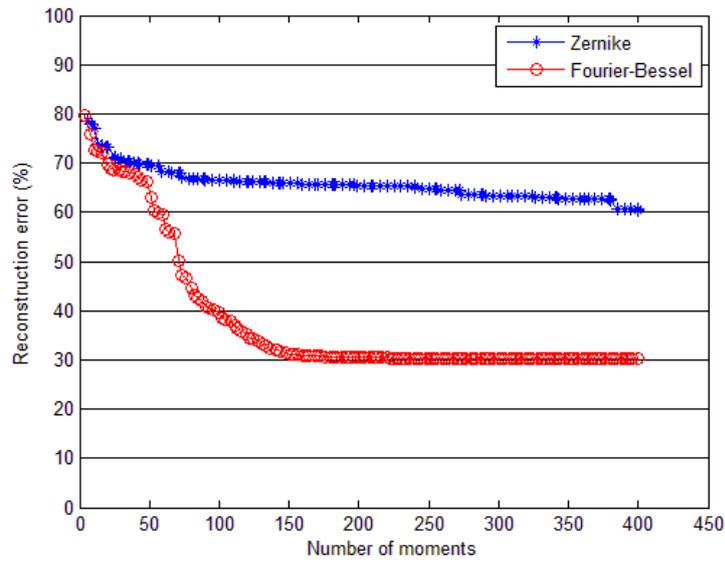

Figure 9 Reconstruction error decrease with increasing number of moments, for Zernike (blue *) and Fourier-Bessel (red o) expansions.

Table 1 Confusion matrix on Gram+/Gram-/Yeasts discrimination at 6h of growth (classification rates in %), using 121 Zernike moments. Database of approximately 1900 scattograms from 15 strains.

|  | Classified as | | |
| --- | --- | --- | --- |
|  | Gram+ | Gram- | Yeasts |
| Gram+ | **92.7** | 3.4 | 3.9 |
| Gram- | 2.6 | **92.4** | 5.0 |
| Yeasts | 0.4 | 2.8 | **96.8** |

Average correct classification rate: 94.7% ± 0.1%*

*standard deviation over ten 10-fold cross-validations

Table 2 Confusion matrix on *E. coli* strain-level discrimination at 6h of growth (classification rates in %), using 121 Zernike moments. Database of 400 scatterograms from 4 strains.

|  | Classified as | | | |
| --- | --- | --- | --- | --- |
|  | ATCC 25922 | ATCC 35421 | ATCC 11775 | ATCC 8739 |
| ATCC 25922 | **96** | 1 | 1 | 2 |
| ATCC 35421 | 7 | **80** | 4 | 9 |
| ATCC 11775 | 6 | 6 | **80** | 8 |
| ATCC 8739 | 5 | 4 | 2 | **89** |

Average correct classification rate: 87.0% ± 0.6%*

* standard deviation over ten 10-fold cross-validations

Table 3 Confusion matrix on *E. coli* strain-level discrimination at 6h of growth (classification rates in %), using 500 Fourier-Bessel moments. Same database as for Table 2.

|  | Classified as | | | |
| --- | --- | --- | --- | --- |
|  | ATCC 25922 | ATCC 35421 | ATCC 11775 | ATCC 8739 |
| ATCC 25922 | **97** | 0 | 0 | 3 |
| ATCC 35421 | 0 | **100** | 0 | 0 |
| ATCC 11775 | 5 | 0 | **87** | 8 |
| ATCC 8739 | 5 | 1 | 5 | **89** |

Average correct classification rate: 93.0% ± 0.7%*

* standard deviation over ten 10-fold cross-validations